\documentclass[aip,pop,reprint,showpacs,preprintnumbers,amsmath,amssymb,a4paper,numerical]{revtex4-1}

\usepackage{layouts}
\usepackage{dcolumn}
\usepackage{csquotes}
\usepackage{bm}

\usepackage{amsmath}
\usepackage{units}

\usepackage{booktabs}
\usepackage{graphicx}
\usepackage{array}
\usepackage{tabularx}

\usepackage[floatwidth=sidefil,capposition=bottom]{floatrow}

\newfloatcommand{ttabboxBC}{table}[\nocapbeside][\FBwidth]

\usepackage{xcolor}     
\definecolor{darkG}{rgb}{0,0.3,0.2}
\usepackage{hyperref}
\hypersetup{
  colorlinks   = true, 
  urlcolor     = blue, 
  linkcolor    = red, 
  citecolor   = darkG 
}

\definecolor{linnumColB}{rgb}{0.2,0.45,0.65}

\usepackage{natbib}

\begin{document}


\title{A neural network approach to kinetic Mie polarimetry for particle size diagnostics in nanodusty plasmas}
\author{Alexander Schmitz}\email{schmitz@physik.uni-kiel.de}
\affiliation{IEAP, Christian-Albrechts-Universit\"at, D-24098 Kiel, Germany\\}
\author{Andreas Petersen}
\affiliation{IEAP, Christian-Albrechts-Universit\"at, D-24098 Kiel, Germany\\}
\author{Franko Greiner} \email{greiner@physik.uni-kiel.de}
\affiliation{IEAP, Christian-Albrechts-Universit\"at, D-24098 Kiel, Germany\\}
\address{Kiel Nano, Surface and Interface Science KiNSIS, Kiel University, Germany}
\date{\today}

\begin{abstract}

The analysis of the size of  nanoparticles is an essential task in plasma technology and dusty plasmas. Light scattering techniques, based on Mie theory, can be used as a non-invasive and in-situ diagnostic tool for this purpose. However, the standard back-calculation methods require expertise from the user. To address this, we introduce a neural network that performs the same task. We discuss how we set up and trained the network to analyze the size of plasma-grown amorphous carbon nanoparticles (a:C-H) with a refractive index $n$ in the range of real($n$) = $1.4-2.2$ and imag($n$) = $0.04i-0.1i$ and a radius of up to several hundred nanometers, depending on the used wavelength. The diagnostic approach is kinetic, which means that the particles need to change in size due to growth or etching. An uncertainty analysis as well as a test with experimental data are presented. Our neural network achieves results that agree with those of prior fitting algorithms while offering higher methodical stability. The model also holds a major advantage in terms of computing speed and automation.

\end{abstract}

\maketitle

\section{INTRODUCTION}


Reactive, particle-generating low-pressure plasmas are widely used in technology\cite{kortshagen2016} and basic research. From reactive components such as acetylene, methane, or silane, particles of up to micrometers in size can grow in the plasma in a multi-stage process. These particles are spherical and monodisperse due to the special properties of the plasma\cite{berndt_2009_some}. 
The investigation of such complex systems often requires 
tracking the particle size over time, for which different in-situ and ex-situ methods exist.
Ex-situ methods use e.g. scanning electron microscopy (SEM) on nanoparticles extracted from the nanodusty plasma\cite{kruger2018,Chutia2021,Asnaz2022}. The invasive extraction itself however is prone to destroying the particle-growing plasma (van-Wetering-hickup\cite{vanWetering2016}), an extraction technique that minimizes the drawback of the particle extraction was presented recently\cite{dworschak2021}. 
Non-invasive, in-situ methods are based on extinction measurements\cite{barbosa2015}, even combined with Microwave Cavity Resonance Spectroscopy\cite{donders2022}, or on the analysis of the polarization state of light scattered by the  particles\cite{greiner_2018_nano}. 
These approaches, however, often rely on the knowledge of the particle's optical properties, most notably the refractive index\cite{Onofri2011}.

From the change in polarization during the growth process, not only can their time-dependent radius course be determined by using the so-called kinetic Mie polarimetry, but this technique also provides the complex refractive index. This is also needed to estimate the absolute particle density from extinction measurements.

Accurately measuring the full polarisation state of light scattered by nanoparticles is a challenge, as the scattered light may include an unpolarized component from stray light and the ambient plasma. A standard method without spatial resolution uses a rotating compensator polarimeter\cite{hauge1975} to measure the polarisation state at a single point in a dusty plasma \cite{hayashi_1994_analysis, gebauer_2003_insitu, groth_2015_kinetic}. The inherently under-determined problem of calculating the particle's refractive index and radius from the observed polarization (the kinetic Mie problem) is addressed by matching the changing polarization state with Mie theory via a fit. As the particles can be considered homogeneous spheres, standard, single scattering Mie theory can be used for the analysis\cite{bohren}. Nevertheless, the strong non-linearity of the problem leads to instabilities in the fitting procedure which, as Hollenstein pointed out\cite{hollenstein_1994_diagnostic}, heavily relies on the expertise of a human operator. In the last years imaging Mie polarimetry is used more and more to investigate also the spatial evolution of dust growth\cite{greiner2012,Groth_2019}. Hence, new challenges arise such as the requirement for stable, automated and fast computation. Current algorithmic methods like CRAS-Mie\cite{groth_2015_kinetic} are not fully up to the task. However, neural networks are well-known for their capability to address such non-linear mapping problems, due to their inherent non-linear structure. We present a new deep-learning approach to the mapping problem via our High-Efficiency Refractive index MappIng NEural network\cite{HERMiNe} {(HERMiNe)}. We show that the accuracy of the resultant refractive index, as well as computing speed, has been significantly increased, compared to the fitting procedure. This paves the path for future data-intensive, real-time imaging of the particle’s growth dynamics in nanodusty plasmas.

In the following section, we give a short introduction to Mie scattering, polarization measurement and the overall problem. In section \ref{sec:nna} the neural network approach is described with its corresponding architecture, training as well as a Monte Carlo based error estimation.
Section \ref{sec:ExperimentalAppl} further shows the network's application to experimental data and draws a performance comparison to the CRAS-Mie fitting method. Finally, the conclusion is given with an outlook on future work.


\section{The Kinetic Mie Problem\label{sec:setup}} 
\label{sec:KineticMieProblem}
\begin{figure*}
   \centering
    \includegraphics{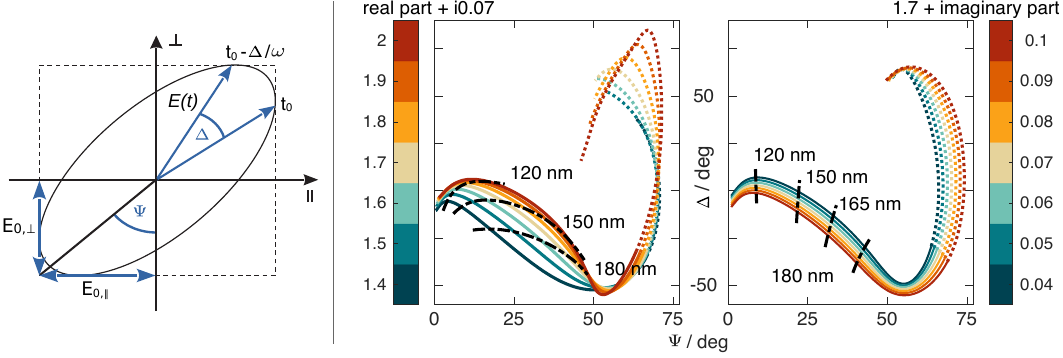}
    \caption{\textit{Left:} Definition of the ellipsometric angles $\Psi$ and $\Delta$ on the polarization ellipse. The oscillating electric field vector $E(t)$ has a component parallel ($\parallel$) and perpendicular ($\perp$) to the scattering plane. The phase difference between those is depicted by $\Delta$, whereas the ratio of the corresponding amplitudes ($E_{0,\parallel},E_{0,\perp}$) gives $\tan(\Psi)$. \\
\textit{Right:} Exemplary $\Delta(\Psi)$ curves for a variety of color-coded refractive indices $n$, calculated over a range of particle sizes $a$ from Mie theory using a wavelength of $\unit[663]{nm}$. For each curve either the imaginary or real part of $n$ is held constant. The dashed black lines represent hypersurfaces of constant $a$. The curve sections plotted with continuous lines indicate the training domain of our network. \the\columnwidth
}
    \label{fig:Psi_Delta_curves}
\end{figure*}

\subsection{Mie Scattering}

Once a ray of light encounters a nanoparticle floating in the plasma environment it gets scattered.
Considering the particle to be approximately spherical\cite{groth_2015_kinetic}, the whole scattering process is well described via Mie-theory\cite{mie1908a} and solely dependent on the 
particle's complex refractive index $n$ and the so-called size parameter $x$. This is the ratio of the particle radius $a$ and the light's wavelength $\lambda$, $x={2\pi a}/{\lambda}$. 
A complete description of fully as well as partially polarized light and their respective measurement instructions are given by the Stokes-Mueller formalism, which we briefly introduce now. A full treatise can be found e.g. in \cite{bohren,collett2012}.
In the Stokes-Mueller formalism, an arbitrary plane electromagnetic wave is characterized by four distinct, intensity-valued parameters $I$, $Q$, $U$ and $V$. Component I measures the total intensity, composed of the polarized and unpolarized parts. 
The other three each describe the predominance of a certain polarization state. Together, they create the Stokes Vector
\begin{equation} \label{eq:Steoksvector}
    S = \begin{pmatrix} I \\ Q \\ U \\ V \end{pmatrix} = 
    \begin{pmatrix}  I_\parallel + I_\perp  \\ I_\parallel - I_\perp \\ I_{\pi/4} -I_{-\pi/4} \\ I_R -I_L \end{pmatrix}  \,,
\end{equation}
where $I_\parallel$, $I_\perp$, $I_{\pi/4}$ and $I_{-\pi/4}$ stand for the linear polarized intensities at geometric angles of 0, 90, 45 and 135 degrees with respect to the scattering plane. The fractions of purely left and right circular polarized intensity are given by $I_L$ and $I_R$ respectively. With this, the polarization state is fully determined by a set of measurable intensities.
%
The degree of polarization is further defined as
\begin{equation}
    \mathrm{DoP} = \frac{\sqrt{Q^2+U^2+V^2}}{I} \leq 1\,,
\end{equation}
which equals unity for fully polarized light only. In general, one has to account for an unpolarized fraction from stray light or the ambient plasma glow. In this case Eq.\,\ref{eq:Steoksvector} can be rewritten as a superposition of both fractions:
\begin{align}
S =& S_{u}+S_p = (1-\mathrm{DoP}) \begin{pmatrix} I \\ 0 \\ 0 \\ 0 \end{pmatrix} + \begin{pmatrix} \mathrm{DoP} \cdot I \\ Q \\ U \\ V \end{pmatrix}\,,\\
S_p &= \sqrt{Q^2+U^2+V^2} \begin{pmatrix} 1 \\ -\cos{(2\Psi)} \\ \sin{(2\Psi)}\cos{(\Delta)} \\ -\sin{(2\Psi)}\sin{(\Delta)} \end{pmatrix} \,.
\end{align}
This separation allows us to use the Stokes vector for polarimetric calculations, even when applied to partially polarized light.
The last equation deploys the identities\cite{collett2012} for the ellipsometric angles $\Psi$ and $\Delta$, which are defined by the wave's time-averaged polarization ellipse according to Fig.\,\ref{fig:Psi_Delta_curves}. Hence, $\Psi$ represents the tangent ratio of the field amplitudes, while $\Delta$ is the phase difference between both field components.
\\

For the sake of completeness, it should be mentioned, that the Stokes-Mueller formalism allows for a complete algebraic description of light passing through any optical setup. This is done by left-side multiplications of the corresponding Mueller matrices of each optical element to the Stokes vector representing the incident light. The Mueller matrices of standard components like polarizers or wave plates and even light scattering by nanoparticles  can be found in \cite{collett2012} or other textbooks.

\subsection{Kinetic Inverse Mapping}
If the incident light beam is well-characterised and the polarization state of emergent light is measured, the change in the light's polarization state, denoted by $\Psi$ and $\Delta$, is only dependent on the scattering process and thus contains all relevant physical information about the scattering target. Strictly speaking, for a known wavelength $\lambda$, we are looking for the inverse map
\begin{equation}
(\Psi, \Delta) \longmapsto (a, n)\,.
\end{equation}
It is clearly visible, that this map is under-determined, since the refractive index $n$ is a complex number. 
A feasible solution to this is the kinetic approach. Instead of treating every ($\Psi,\Delta$)-point individually, the time evolution of the target is exploited. When considering a varying (e.g. increasing) particle radius, unique curves, represented as $\Delta(\Psi(a))$, form for every possible refractive index. 
Fig. \ref{fig:Psi_Delta_curves} shows such a collection of possible $\Delta(\Psi(a))$ curves from growing particles of different refractive indices, as observable in our experiment (see Sec.\,\ref{sec:setup}). 
Here the ambiguity and non-linearity of the Mie problem in $n$ is clearly visible. 

The inverse map can be computed by means of a kinetic fitting algorithm, as presented by Groth et al\cite{groth_2015_kinetic}: Via an iterative least-square method, the curve's course is fitted to a theoretical model, assuming a monodisperse size distribution at each moment and the resultant refractive index to be constant in time. When the refractive index is known, each point of the $\Delta(\Psi(a))$ curve is unambiguously associated with a radius $a$.
\\

While this fitting method yields particle sizes in good agreement with ex situ measurements\cite{Groth_2019,dworschak2021}, current numerical procedures need a degree of judgment expertise, for they tend to be sensitive on start parameters, particularly when performed on shorter curve segments. 

%
%
\section{Neural Network Approach}
\label{sec:nna}

Instead of fitting, the underlying problem of curve shape regression can also be approached by means of machine learning. Here, the complex relation between a $\Delta(\Psi)$ curve and its specific refractive index is encoded into a deterministic, neuronal structure, rather than transferred to a fitting algorithm via a model function. 
This method not only enables faster data evaluation, but also a potentially better problem convergence, due to a network's well known inherent capability to reproduce nonlinear dependencies. The challenge here is, that careful and extensive training is needed, to find the correct network weights.

We present our High-Efficiency Refractive Index MappIng NEural network\cite{HERMiNe} ({HERMiNe}) as a suitable solution for the inverse kinetic Mie problem.

\subsection{Neural Network and Training}
\begin{figure*}
\begin{floatrow}
\ffigbox[\FBwidth]{%
\includegraphics[width=12.7cm]{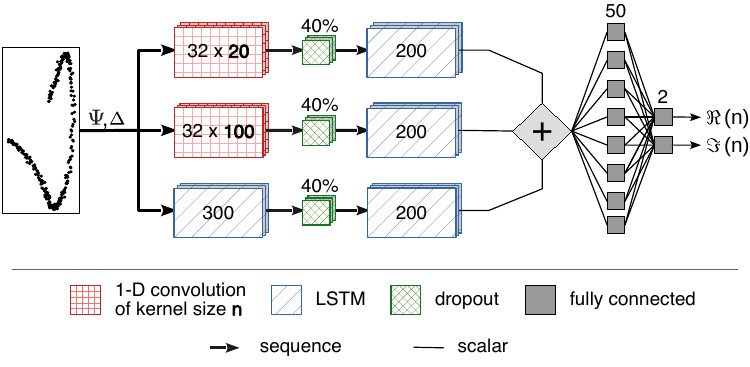} 
}{\caption{Network Layer Graph. The network consists of 1D convolutional, long short-term memory (LSTM), dropout and fully connected layers of depicted unit quantity. The kernel size of the convolutions is shown as bold numbers, the dropout layer's probability is given in percentage. As a sequence, the $\Delta(\Psi)$ curve is first fed through three differing branches. The last LSTM layer group outputs one scalar value per unit, which are added element-wise along the branches and further reduced afterwards. The final outputs are the predicted components of the complex refractive index $n$, which, in turn, unambiguously define a $\Delta(\Psi(a))$ curve as reference.}\label{fig:layergraph}}
\killfloatstyle\ttabboxBC[\Xhsize]{%
\renewcommand{\arraystretch}{1.2}
\begin{tabularx}{\Xhsize}{>{\raggedright\arraybackslash}X >{\raggedleft\arraybackslash}X}
\toprule
  \multicolumn{2}{c}{Network domain} \\ \midrule \hline
  Real part & $1.4-2.2$ \\
  Imaginary part & $0.04i-0.1i$ \\
   Angle of observation & $\unit[90]{deg}$\\
      &  \\
   \multicolumn{2}{c}{Training parameters}\\ \hline
  Number of samples & $247500$ \\
  Mini-batch size & $50-100$ \\
  Initial learning rate & $0.001$ \\
  Learning method & Adam \\
    \end{tabularx}
}{\caption{Network parameter domain and training conditions.}%
\label{tab:trainingParams}}
\end{floatrow}
\end{figure*}

All network architecture, training and validation were realized using the \textsc{Matlab} 2022a \textsl{Deep learning toolbox}\cite{MatlabDL}.
The network structure, as displayed in Fig.\,\ref{fig:layergraph}, 
has shown to be the most promising of several approaches and was further refined and fine-tuned.

\subsubsection{Network Architecture}

We choose to take the ($\Psi,\Delta$)-input as 2D series data rather than interpreting the curve as an image for convolutional and pooling processing. With this, model generalization should be achieved more easily with less training, since image orientation, centering and corresponding preprocessing of measured data can be disregarded.
We make use of a many-to-one approach to predict a scalar, complex refractive index from a sequential curve input. The radii of the input ($\Psi,\Delta$)-sequence can then easily be obtained by associating each measuring point to the nearest point on the now-identified reference curve, given by the predicted $n$.
In theory, a sequence-to-sequence network could directly predict the time dependent radius from a curve input. This task, however, introduces more degrees of freedom and therefore requires substantially more training effort, while not offering a benefit to our method.

According to the universal approximation theorem, almost every network should be capable of correctly reproducing any problem, given enough neurons and infinite training time. Finding a network of appropriate type and structure however drastically reduces computing costs and is therefore a vital task. 

An ordinary Multilayer Perceptron (MLP), which is easy to handle and train, is less suitable for our sequential data, since it only features an input of fixed length.
When considering time series data of varying lengths, the MLP heavily relies on zero padding. For our problem, we find that this approach shows poor training convergence.
A Recurrent Neural Network (RNN) on the other hand uses recursion to process flexible input sizes. It's structure allows to link together nearby time steps throughout the series and therefore to pass on context about the curve's previous shape. Though, this information degrades rapidly over the course of the data series, what makes it difficult to train long term dependencies. This vanishing gradient problem is an inherent structural property of RNNs.
A solution to this are Long Short-Term Memory (LSTM) networks, as introduced by Hochreiter\cite{hochreiter1997}, which we make use of in our network design, in combination with 1D convolutional layers. 
\\

The architecture is divided into three parallel branches, containing a preprocessing, dropout and LSTM layer. The branches are thereafter combined to form an output.

Two branches feed the $(\Psi,\Delta)$-sequence through a 1D convolutional layer for preprocessing, which is used to extract differential features. The maximum possible distance for cross-related information along the curve is dependent on the kernel size, which is why we make use of two different kernels in each of the network's branches. One for mid- (size $100$) and the other for short-distance (size $20$) correlation. In the third branch, the leading layer is a LSTM ($300$ units), to allow for a self-trained, arbitrary relation range.

Each branch is continued with a dropout layer to randomly discard an output node from the previous layer during training with a probability of $\unit[40]{\%}$. This statistical disconnection offers a cheap and effective regularization technique to battle overfitting and enhance model generalization by simulating a continuously changing network structure. 

The preprocessed sequences are transferred over to many-to-one LSTM layers respectively ($200$ units each). The resultant scalar outputs from the units of all branches are added element-wise to a $200\times1$ vector and further interlinked by two fully connected (fc) layers ($50$ and $2$ neurons) to produce a final output - the refractive index, which give the reference $\Psi(\Delta)$ curve. 

For all LSTM layers the gate activation functions are standard sigmoid with tanh as state functions. The fc layer's activations were chosen to be linear for this regression problem.
The loss function to be minimised during training is the mean-squared-error of the output refractive index. To compensate for the fact, that the real part of $n$ is a magnitude higher than the imaginary, the value of the latter is weighted by a factor of ten during processing. 

\subsubsection{Network Training}
\begin{figure*} 
%
%
\centering
    \includegraphics{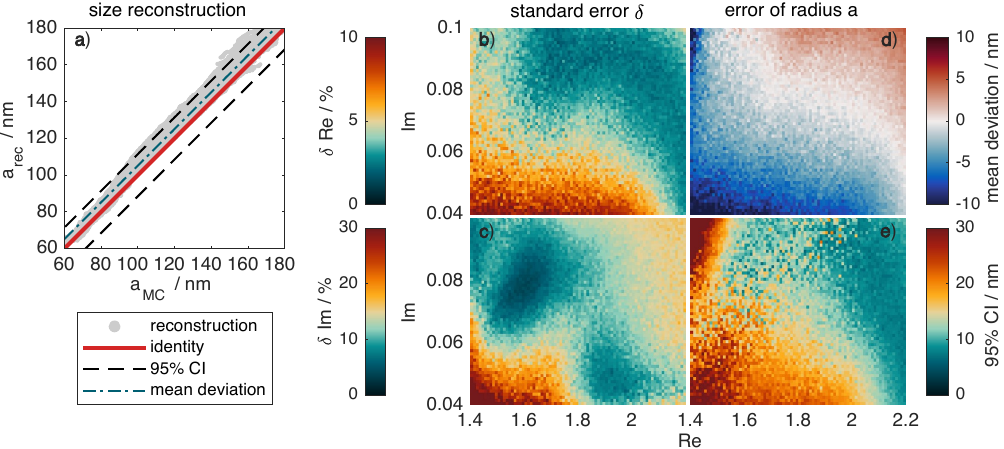}
    \caption{Monte Carlo error estimates for refractive index and particle radius on each point of the trained parameter space. 
    (a) Exemplary analysis of the Monte Carlo method for the refractive index $1.7+i0.07$: The gray dots depict the ensemble of simulated curves as their radius distributions $a_\mathrm{MC}$ and respective reconstructed values $a_\mathrm{rec}$ from the neural network predictions. The continuous red line marks the identity, whereas the black dashed lines mark the $\unit[95]{\%}$ confidence interval around it. The mean deviation of the ensemble is indicated by the blue dot and dash line.
    (b) Standard error $\delta$ of the mean for real (Re) and (c) imaginary (Im) part of refractive index prediction. (d) Mean ensemble deviation of reconstructed particle radius. (e) Width of the $\unit[95]{\%}$ confidence interval (CI) of the reconstructions.}
    \label{fig:Error_Estimation}
\end{figure*}
Unfortunately, there is no large set of measured $\Delta(\Psi)$ curves available for which the actual refractive indices have been verified, e.g. via scanning electron microscopy\cite{groth_2015_kinetic}. In order to train the network we resort to the use of synthetic, thus simulated, data, that closely resemble real measurements. 
This method is commonly and successfully employed in applications where training data is scarce or costly and provides greater control over edge cases and statistical data biases. 

As such, {HERMiNe} was trained using $247\,500$ artificial $\Psi$-$\Delta$ curves of varying length computed directly from Mie theory, where the scattering targets are assumed to be spherical. 
The trained section of the curve is highlighted in Fig.\,\ref{fig:Psi_Delta_curves} and encompasses radii up to several hundred nanometers, depending on the refractive index and wavelength.

Table\,\ref{tab:trainingParams} provides an overview of the network parameter domain and training conditions.
Taking previous findings\cite{greiner_2018_nano} into consideration, the incorporated range of refractive indices covers $1.4-2.2$ for the real and $0.04i-0.1i$ for the imaginary part. 
The overall scattering conditions were chosen to reflect our experimental measurement setup, such as employing linear $\unit[45]{deg}$ polarized incident light ($S = (1, 0, 1, 0)^\mathrm{T}$) and observation under $\unit[90]{deg}$ with respect to the scattering plane. Also, Gaussian noise of $\sigma = \unit[1]{deg}$ was added to the data to better resemble real data from our measurement device (see sec.\,\ref{sec:setup}).
To facilitate model generalization, robustness and prevent data biases, we 
introduce diversity into our simulated growth patterns e.g. by varying the resolution (point density) and sample rate. 

The actual training was performed in multiple sessions, using the Adam (Adaptive Moment Estimation) algorithm for weight updates with a declining learning rate that started at $0.001$. After each session, we evaluated the networks performance, and based on the error detected, generated new data for further fine-tuning.

\subsection{Model Validation}

Due to the nature of neural networks, it is not feasible to specify a distinct uncertainty, unlike with e.g. the residuum of a fitting algorithm.
A network can be seen as a black box that consists of many highly nonlinear operations to generate a response to a specific input. Its output precision may vary unforeseeable for different input conditions, such as refractive indices, curve sections or noise levels. It is important to note, that the network will always give the exact same result for identical inputs.

Statistical methods such as Bootstrapping\cite{Numrecipes} or Monte Carlo simulations can be used in order to properly validate model convergence to the training data, its precision and ability to generalize.

\subsubsection{Monte Carlo Error Estimation}

To obtain an average error estimate for the network, we utilize a Monte Carlo method, described in the following. For each point in the network's parameter plane, which is defined by the real and imaginary part of $n$, an ensemble of $100$ different theoretical $\Delta(\Psi)$ curve sections of varying lengths is created. Normally distributed noise of $\sigma=\unit[1]{deg}$ is then added to the data, what resembles the typical statistical spread of our measurement device.
These data sets are fed into the network to retrieve predictions for refractive indices and, from that, to recalculate the corresponding radius progressions. We are interested in the standard error of the refractive index prediction, as well as an error estimate for the reconstructed radii. For this, we use mean deviation compared to the original data and the corresponding $\unit[95]{\%}$ confidence interval (CI).

For better clarification, Fig.\,\ref{fig:Error_Estimation}, (a) illustrates the meaning of CI and mean deviation, and how they are derived. 
For one exemplary $n$-parameter point, all $100$ ensemble $\Delta(\Psi)(n)$ curves are shown stacked in grey as radius reconstruction $a_{rec}$, based on the network predictions, against original radius $a_{MC}$. The identity, marked in red, would resemble a perfect fit. The mean deviation is given by the average distance of all reconstructed radii to this identity, indicated in blue. The dotted lines correspond to the $\unit[95]{\%}$ confidence interval around the identity.
\\

The results are displayed in Fig.\,\ref{fig:Error_Estimation} and provide an error estimate, resolved over the parameter plane $n$. The map shows the refractive index standard error of the mean of each ensemble, separately for real and imaginary part. It is visible that the average deviation is highest for curves where both the real and imaginary part of $n$ are low. The error of the imaginary part is typically larger than that of the real part. This is likely caused by its lesser impact on the curve's shape for the chosen number range of $n$, which, in turn, fortunately also reduces the sensitivity of the radius to uncertainties in the imaginary part (cf. lines of constant radius in Fig.\,\ref{fig:Psi_Delta_curves}).
The Fig.\,\ref{fig:Error_Estimation}, (d) shows the previously mentioned mean deviation of the reconstructed radii estimation compared to the original values for every ensemble. Positive numbers indicate an overestimation of particle sizes by our method, negative values indicate underestimation. Again, the largest error lies in the region of low refractive index, both real and imaginary. 
The corresponding width of the ensemble's $\unit[95]{\%}$ confidence interval (CI) for the radius shows a similar distribution in Fig.\,\ref{fig:Error_Estimation}, (e). 

Note that differences in the error patterns from Fig.\,\ref{fig:Error_Estimation}, (d), (e) compared to those seen in b), c) result from the non-uniform sensitivity of the radius to changes in the refractive index (cf. Fig.\,\ref{fig:Psi_Delta_curves}). Therefore, the error in the radius estimation is more pronounced in some regions than expected from the refractive index.

\begin{figure}
    \centering
    \includegraphics[width =\columnwidth]{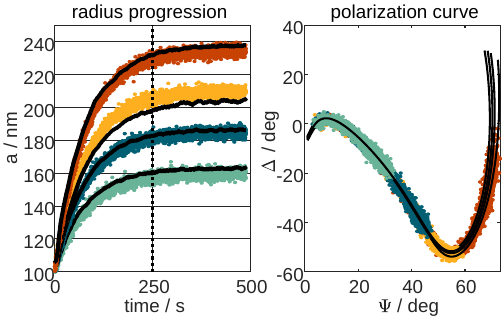}
    \caption{Network's Response to simulated non-linear particle growth. \textit{Left:} Saturating progressions of particle radiusF over time. The color-coded scatter plots show the original simulated courses. Black lines represent the smoothed recalculation from the network's prediction. The part of the data sequence which was fed into the network ends with the vertical mark. \textit{Right:} $\Delta(\Psi)$ curves for each course calculated from Mie theory with $1.7+i0.08$ and $\unit[1]{deg}$ Gaussian noise, plotted on top of each other. The colors match those on the right side. As the final particle size gets larger and larger, the $\Delta(\Psi)$-series progresses further. The black continuous lines show the refractive indices estimated by the network, depicted as what the corresponding $\Delta(\Psi)$ curves would look like.}
    \label{fig:NonLinGrowth}
\end{figure}

\subsubsection{Non Linear Particle Growth}
Besides the Monte Carlo based estimation of the average error statistics discussed previously, the behavior of the network in special cases is important. 
In particular, for our laboratory applications, it is of special interest to examine particle growth progresses that are non-uniform, where the size initially grows linearly in time but eventually reaches a saturation point, causing the radius to approach a limiting value. This scenario is relevant for the production of nanoparticles with a specific target size.

To address the network's performance on this question, different simulated $\Delta(\Psi)$ curves of such saturating particle size progressions are evaluated and their radii reconstruction, based on the networks prediction, compared to the original.
In Fig.\,\ref{fig:NonLinGrowth} four such curves and their corresponding analysis are shown as exemplary representatives, all for a refractive index of $1.7 + 0.08i$. The sequences are evaluated up to the time step of $\unit[250]{s}$. Supplying time steps beyond this can decrease the network's performance, as the saturated regime then dominates the sequence.
As demonstrated by the continuous black lines in the display, the network predictions accomplish high congruence with the predetermined curves, with only little scatter. The maximum deviation lies at $\unit[4]{\%}$ for the real part and at $\unit[18]{\%}$ for the imaginary part. Similar findings were obtained for other values of $n$ and radius progressions, where some evaluations show less and others greater deviation as in this example. Overall, the analysis confirms that our method is well capable of processing non-linear growth rates. It also demonstrates the stability of the network's predictions, as it provides a similar result when confronted with curves of the same refractive index but different conditions like growth rate and progression.

\section{Experimental Application}
\label{sec:ExperimentalAppl}
The next step is to let {HERMiNe} evaluate real-world data. Although the network
shows robustness so far, experimentally acquired data lacks the pre-specified, well-controlled statistics of theoretical data sets used for training. Discrepancies in noise width and statistics or systematic errors from the measurement devices and setup - such as a slightly different observation angle - can negatively impact the network's predictoin. The result will show, if the training sufficiently reproduced the experimental applications.

\subsection{Experimental Setup} \label{sec:setup}
\begin{figure}
\centering
\includegraphics[width=\columnwidth]{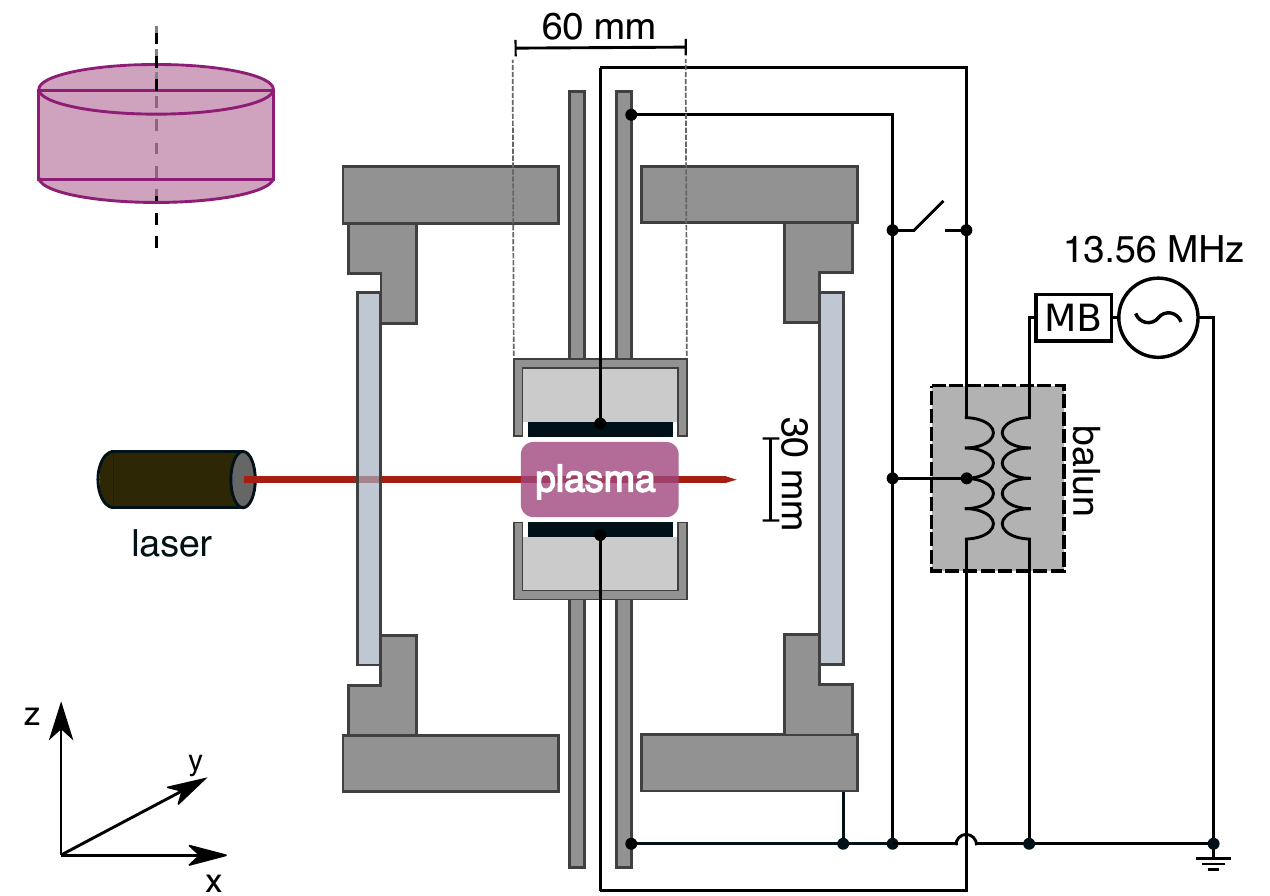}
\caption{Side view sketch of the experimental setup. The plasma is created between two symmetrically powered electrodes. The circuit consists of a $\unit[13.56]{MHz}$ rf-generator, an impedance matching network (MB) and a balancing unit (balun). The vacuum chamber, electrodes and plasma exhibit a symmetric cylindrical configuration.
A vertically orientated $\unit[662.6]{nm}$ laser with a linear $\unit[45]{deg}$ polarization is used as illumination. The measuring instrument is mounted orthogonally in y-direction.}
\label{fig:setup}
\end{figure}

\begin{figure}
\centering
\includegraphics[width=\columnwidth]{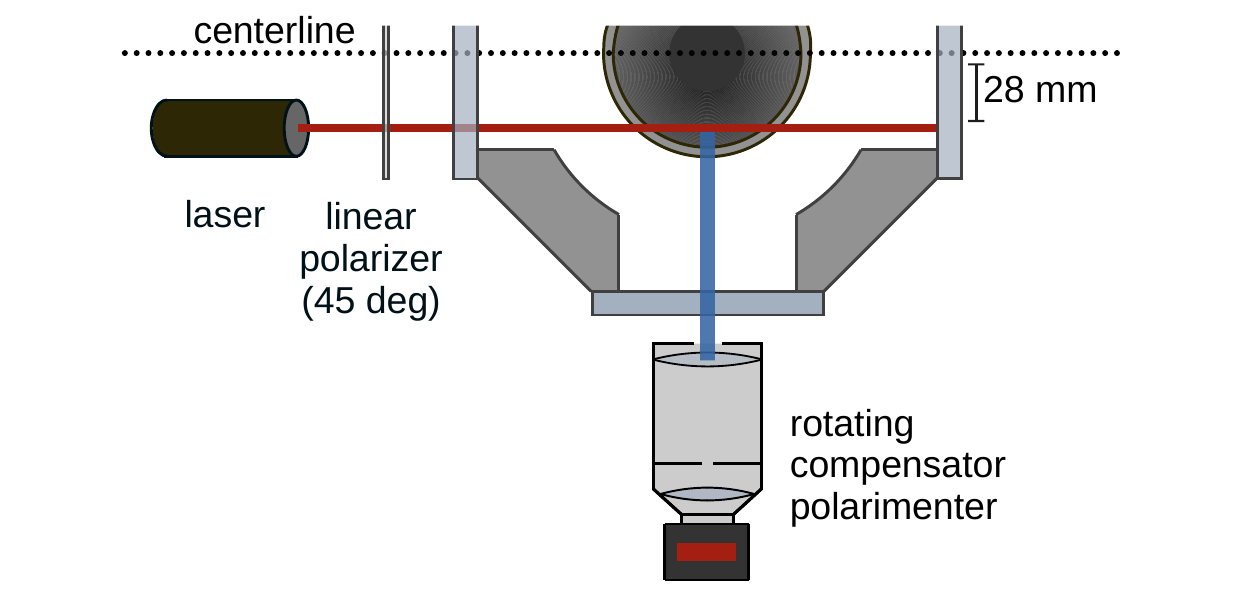}
\caption{Top view of the laser and measurement arrangement. The linear $\unit[45]{deg}$ polarized laser of $\unit[662.6]{nm}$ wavelength illuminates the plasma $\unit[28]{mm}$ off center. The rotating compensator polarimeter is mounted perpendicularly to the laser beam, measuring the scattered light emerging from a fixed point, as indicated in blue.}
\label{fig:setupTop}
\end{figure}

Our measurement setup consists of a capacitively coupled parallel plate argon discharge, which is rf-driven at $\unit[13.56]{MHz}$ and $\unit[8]{W}$. For more details, a cross-sectional view is provided in Fig.\,\ref{fig:setup}, a top view is presented in Fig.\,\ref{fig:setupTop}. The cylindrical electrodes have a diameter of $\unit[60]{mm}$ and are spaced at $\unit[30]{mm}$ distance.
To ensure a symmetrical operation mode with reference to the electrically grounded stainless steel vacuum chamber, a balancing unit (balun) is used. An impedance matching network is necessary to maximize power transfer to the plasma and reduce rf-reflection.

The gas flow, directed top down at a constant rate of $\unit[8]{sccm}$, is a mixture of our work gas argon and $\unit[20]{\%}$ acetylene as reactive component to grow the nanoparticles. The working pressure is set to $\unit[21]{Pa}$. 
A vertically orientated, red ($\unit[662.6]{nm}$) laser, linearly polarized at $\unit[45]{deg}$ with respect to the horizontal plane, illuminates the arising nanoparticle cloud, which is known to exhibit cylindrical symmetry \cite{greiner_2018_nano}.

During the experiment, a rotating compensator polarimeter (RCP), mounted perpendicularly in negative y-direction, monitors the polarization state of the light scattered by the growing nanoparticles.
The values of $\Psi$ and $\Delta$ are extracted from the harmonics of a Fourier analysis, as described by Hauge and Dill\cite{hauge1975}. The average measuring inaccuracy of the device can be well approximated by Gaussian noise of $\sigma=\unit[1]{deg}$ for both measurands. For very small particles, though, the signal-to-noise ratio imposes a detection limit.

\subsection{Results}
A total of eleven independent nanoparticle growth cycles along with the corresponding $\Delta(\Psi)$ polarization curves were recorded under similar experimental conditions on different days. Prior investigations on the discharge have demonstrated that the observed system is well reproducible\cite{petersen2022}. Therefore the curves can be considered similar and their refractive indices are comparable within the ensemble.
This data set is used to verify the accuracy of the neural network's predictions on real-world measurement data.

The CRAS-Mie fitting method\cite{groth_2015_kinetic} is an established\cite{Groth_2019,dworschak2021,petersen2022,greiner_2018_nano} technique for obtaining the refractive index of a polarization curve. Its capabilities has been validated by ex-situ particle size measurements using SEM\cite{groth_2015_kinetic} and AFM\cite{dworschak2021}. 
In this work, it is used to provide an approved, mean reference for the refractive index of the data set. 
As mentioned before, the major downside of the CRAS-Mie method is its sensitivity to the start values and the choices of the corresponding operator, which leads to spread in the resultant values or even abortion of the algorithm.

\begin{figure*}
    \centering
    \includegraphics{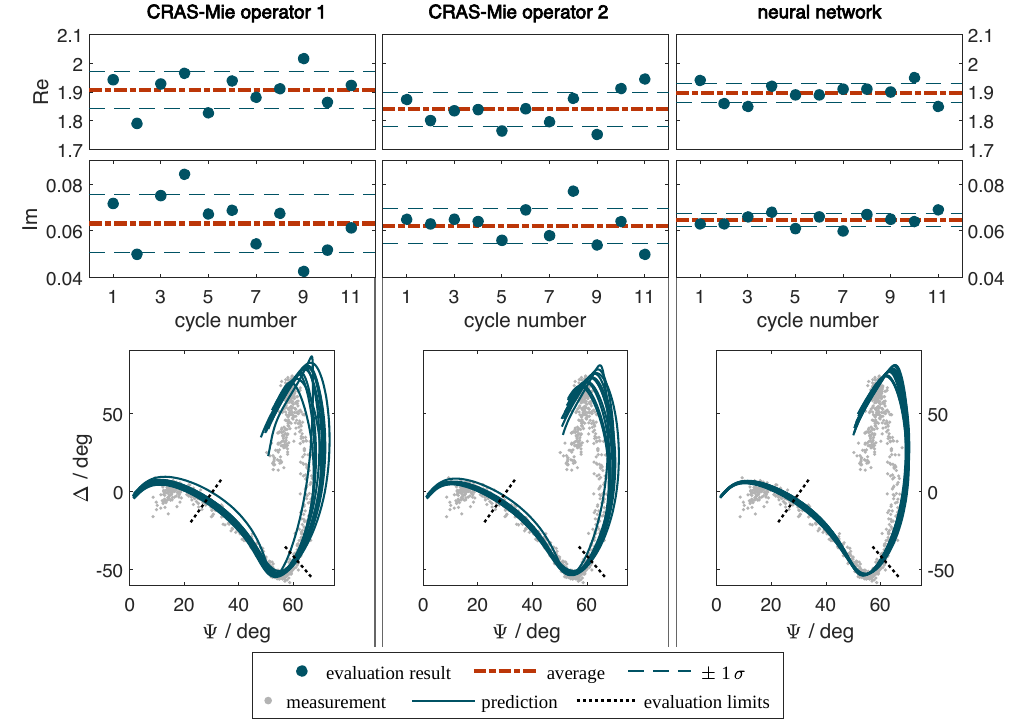}
    \caption{Columnwise comparison of the refractive indices $n$, produced by our network and two different human operators using the CRAS-Mie fitting method. Each is performed on the same measured data set of eleven growth cycles. The scales are identical for all three columns.
    The dots depict the respective results as real and imaginary part of $n$.
    Arithmetic average and $\pm 1\sigma$ range are indicated by dashed and dash-dotted lines.
    The lower part shows the respective eleven resulting refractive indices as $\Delta(\Psi)$-solutions with solid lines. In the background, all data points of one representative growth cycle are depicted as grey dots. The grouping of the other data sets is similar enough, so the single example gives a good impression.
    Dotted lines roughly indicate the section of the measurement data which was used for the evaluations. Outside this region of interest, the particles either approach the detection limit of the sensor or are suspected to undergo a change in the refractive index\cite{kobus2022} or show multi-scattering\cite{kirchschlager2017}, resulting in a deviation to our model. The visible scatter of the data can be well approximated by Gaussian noise with $\sigma =\unit[1]{deg}$.
    }
    \label{fig:OperatorVergleich}
\end{figure*}

In the upper part of Fig.\,\ref{fig:OperatorVergleich}, the network's refractive index prediction for the measurement data is shown in comparison to the same evaluation done by two different human operators, both using CRAS-Mie.
For all analysis, only the central curve section is used, as indicated in the lower part of the figure. 
Outside this region of interest, unwanted effects play a role:
For very small particle radii, the RCP approaches its detection limit, when stray light and noise superimpose the decreasing polarization signal. 
As the particles grow too large, they are suspected to undergo a change in the refractive index\cite{kobus2022} or show multi-scattering\cite{kirchschlager2017}, which is an object of current research.

The analysis results show, that the refractive indices obtained by the three methods are consistent with a similar average outcome. The corresponding reconstructed $\Delta(\Psi)$ curves (solid lines in the lower part of Fig.\,\ref{fig:OperatorVergleich}) are all in good agreement with the evaluated measurement data. 
The prediction error of the network, thus the mean deviation from the physical value, is not directly accessible. However, the comparison of the result with CRAS-Mie, the measurement, and the error-chart in Figure\,\ref{fig:Error_Estimation} indicates a good performance.

There is a significant difference in the spread introduced by each method, again highlighting the influence of the operator's experience and judgement on the result's quality.
The network features significantly less variance for both, the real and imaginary part of the refractive index and removes the human element as a factor, thus offering easier access and a more reproducible and comparable evaluation procedure. It also offers a robust method for the automated evaluation of a high number of curves, as necessary for imaging techniques.

Computation effort shows considerable benefit, as the network reduces the necessary time by an average factor of $7$ over the fit, measured on a standard NVIDIA GTX 1060 graphics card ($\unit[90]{ms}$ compared to $\unit[620]{ms}$). This might be considerably increased when using more powerful hardware.



\section{Conclusion and Outlook\label{sec:conclus}}
We presented a new sequence regression network, using a hybrid convolutional and LSTM structure, to
obtain the radius and refractive index of nanoparticles from their size-dependent polarization signal.
The study demonstrates the robustness of the network's prediction for a range of different scenarios, including non-linear particle growth, and shows that it provides accurate estimates of particle sizes in a wide range of applications.

Comparing the network's performance with the ex-situ verified, non-linear CRAS-Mie fitting technique, we found that the network produces similar results but exhibits significantly reduced scatter. Furthermore, our new method is not dependent on well-chosen start values and the operator's experience, both of which are crucial factors for the fitting algorithm. Our network thus offers a more comparable method for refractive index prediction and enables the robust automated analysis of $\Delta(\Psi)$ curves in imaging polarimetry applications.

The network's advantage in computation speed over the fit facilitates the application of spatially resolved imaging measurements of nanoparticle clouds, which are currently under development. The application of better hardware could even enable real-time monitoring of such systems.

Future improvements will most likely include transfer training to access evaluation of the high-slope section of the polarization curve. The training in this regime holds particular challenges due to the suspected change in the refractive index.

\section*{Acknowledgements}
The present investigations were financially supported by the
Deutsche Forschungsgemeinschaft (DFG) within the project {GR 1608/8-1, project number 418187010}. Thanks go to our colleagues and the indispensable technical staff at the institute.

\section*{References}
\bibliographystyle{unsrtnat}
\bibliography{Literature}

\end{document}